\documentclass[twocolumn,pre,superscriptaddress,bibnotes,notitlepage,nofootinbib,eqsecnum]{revtex4-2}

\usepackage{CJK}

\usepackage{graphicx}

\usepackage{dcolumn}

\usepackage{amsmath}

\usepackage{bm}
\usepackage[normalem]{ulem}
\usepackage{color}
\usepackage{amsmath}
\usepackage{enumitem}

\setlist[itemize]{noitemsep,leftmargin=*}
\usepackage{amssymb}
\usepackage{lipsum}
\usepackage{bm}
\usepackage{xcolor}
\usepackage{graphicx}
\usepackage{verbatim}
\usepackage[T1]{fontenc}
\usepackage{hyperref}
\usepackage{epstopdf}

\newcommand{\bq}{\mathbf{q}}

\newcommand{\beq}{\begin{equation}}
\newcommand{\eeq}{\end{equation}}
\newcommand{\beqn}{\begin{eqnarray}}
\newcommand{\eeqn}{\end{eqnarray}}

\newcommand{\bew}{\begin{widetext}}
\newcommand{\ew}{\end{widetext}}

\begin{document}
\title{
{Response to the comment on The inconvenient truth about flocks by Chat\'e and Solon}}
\author{Leiming Chen}
\email{leiming@cumt.edu.cn}
\affiliation{School of Material Science and Physics, China University of Mining and Technology, Xuzhou Jiangsu, 221116, P. R. China}
\author{Patrick Jentsch}
\email{patrick.jentsch@embl.de}
\affiliation{Cell Biology and Biophysics Unit, European Molecular Biology Laboratory, 
Meyerhofstra\ss e1, 69117 Heidelberg, Germany}
\author{Chiu Fan Lee}
\email{c.lee@imperial.ac.uk}
\affiliation{Department of Bioengineering, Imperial College London, South Kensington Campus, London SW7 2AZ, U.K.}
\author{Ananyo Maitra}
\email{nyomaitra07@gmail.com}
\affiliation{{Laboratoire de Physique Th\'eorique et Mod\'elisation, CNRS UMR 8089, CY Cergy Paris Universit\'e, F-95032 Cergy-Pontoise Cedex, France}}
\affiliation{Laboratoire Jean Perrin, Sorbonne Universit\'{e} and CNRS, F-75005, Paris, France}
\author{Sriram Ramaswamy}
\email{sriram@iisc.ac.in}
\affiliation{Centre for Condensed Matter Theory, Department of Physics, Indian Institute of Science, Bangalore 560 012, India}
\affiliation{International Centre for Theoretical Sciences, Tata Institute of Fundamental Research, Bangalore 560 089 India}
\author{John Toner}
\email{jjt@uoregon.edu}
\affiliation{Department of Physics and Institute of Theoretical
Science, University of Oregon, Eugene, OR $97403^1$}
\affiliation{Max Planck Institute for the Physics of Complex Systems, N\"othnitzer Stra\ss e 38, 01187 Dresden, Germany}
\begin{abstract}
This is our response to the comment \href{https://arxiv.org/abs/2504.13683}{arXiv:2504.13683} posted by Chat\'{e} and Solon in reference to our preprint \href{https://arxiv.org/abs/2503.17064}{arXiv:2503.17064}.
\end{abstract}

\maketitle

{Our position is unchanged: the theoretical analysis in \cite{Solon_Chate} is incorrect, and that their comment \cite{Comment} on our article \cite{Article} in no way refutes this.} 
{Ref.~\cite{Comment} identified two main points of contention. Their issue (ii) is easily disposed of: they do not identify a true invariance of the flock, as they now agree. Their first point (i)} 
{amounts to the claim that Goldstone modes must obey conserved dynamics. As we noted in our original article, and reiterate here, this claim } is wrong, and contradicts the most basic understanding of broken-symmetry hydrodynamics, see, e.g.,
Martin, Parodi, and Pershan \cite{MPP}. 
Solon and Chat\'{e} have not presented any argument that their claim is correct; they merely assert it without proof. 

The validity of the
numerical results of \cite{Solon_Chate} is not an issue we discussed in \cite{Article}, and we will not discuss it here.

We'll now refute the claims of their latest comment \cite{Comment}.

\begin{itemize}
    \item {Ref.~\cite{Comment}, {specifically in the paragraph containing that comment's Eq.~(1)}, reiterates the incorrect argument that rotation symmetry} {\emph{requires} that} {the deterministic part of the dynamics of orientational Goldstone modes in systems that spontaneously break rotation symmetry} 
    {obey} a conservation law. That this argument is incorrect is immediately seen from the 
    counterexample of {nematic liquid crystals} \cite{Article} which are \emph{explicitly} invariant under rotation 
    (which, ipso facto, implies that the ``global direction cannot relax deterministically to another 
    direction''), but in which the deterministic part of the dynamics of the angle field does {\it not} have a 
    conservation law form, as {can be seen from} the extensive literature on the hydrodynamics of nematic liquid crystals.
    
    Contrary to the claim by Solon and Chat\'e in \cite{Comment} that we did not provide any counterexamples involving rotation invariance, this counterexample \emph{was} discussed in our article \cite{Article}. The paragraph containing Eq.~II.6 explicitly discusses the breaking of rotation symmetry without additionally breaking translation symmetry. Further, footnote 3 explicitly discusses passive nematics and polar liquid crystals. Moreover, footnote 1 points out that the KPZ equation may describe the dynamics of a Goldstone mode in an active (and chiral) XY model.

Further, their claims regarding \cite{MPP} are also incorrect. In contrast to their claim, the discussion around Eq.~(2.17) in that article \cite{MPP} only demonstrates that the dynamics of the Goldstone mode of nematic liquid crystals doesn't contain a term $\propto {\bf v}$. The term $A_{ij}^\alpha\partial_jv_i$ can be written as $\partial_j(A_{ij}^\alpha v_i)$ if {\it and only if} $A_{ij}^\alpha$ is {\it not} a function of the fields. Since Ref. \cite{MPP} {focuses on the} \emph{linear} hydrodynamic equations, that is the case they consider. However, {once one goes beyond linear order,} $A_{ij}^\alpha$ {\it does} contain fields; {as a result,} the term $A_{ij}^\alpha\partial_jv_i$ cannot be written as $\partial_j(A_{ij}^\alpha v_i)$ (see any standard reference on nematic liquid crystals such as \cite{Forster} or \cite{Stark} or \cite{deGennes}). 
{In fact, even the coefficient $\lambda$ that appears in Eq.~(2.17) of \cite{MPP}{\textemdash}the very equation that \cite{Comment} claims implies that ``Martin, Parodi and
Pershan~\cite{MPP}... insist on the contrary that, for the breaking of rotational symmetry, the dynamics of the Goldstone mode take the form of [a conservation law]''{\textemdash}is necessarily a function of fields.}  For example, {$\lambda$ depends on the} density through the order parameter amplitude, {diverging at the nematic-isotropic critical point as the inverse of the order parameter amplitude \cite{Forster, Stark}. }

Moreover, as we point out in our article, the advective term ${\bf v}\cdot\nabla\theta$ is not a total divergence.

    {We note in passing that their original article \cite{Solon_Chate} argues for this conservation law \emph{not} on the basis of rotation invariance, but on the grounds that \emph{all} Goldstone modes  must obey a conservation law, which is even more transparently incorrect, as the numerous other counterexamples that we presented in \cite{Article} show.

Furthermore, while  \cite{Solon_Chate} require (incorrectly) that the {\it deterministic} part of their dynamics for the Goldstone mode be a total divergence, {their} noise violates this requirement, {which}  is internally inconsistent. That the{ir} noise does indeed violate this requirement is clear from the fact that its correlations do not vanish as $\bq\to{\bf 0}$.}

{Finally, the explicit argument they present for the conservation-law form in \cite{Comment} is wrong on its face. Notice that if one imposes a constant density gradient along some direction (say $\hat{{\bf x}}$), one \emph{explicitly} breaks the isotropy of space. Therefore, in that state, the angle fluctuations \emph{must} acquire a mass. That is, using their notation from \cite{Comment}, the global direction $\Theta$ (which now measures the deviation of the order parameter away from $\hat{{\bf x}}$) should have the \emph{deterministic} dynamics $\partial_t\Theta\propto-\Theta\langle\partial_x\rho\rangle$, where $\langle\partial_x\rho\rangle$ is the average density gradient. That implies that $\partial_t\theta$ \emph{must} have a term $\propto\theta\partial_x\rho$ {which {\it must not} come from a total derivative}. As we discuss in \cite{Article} and as is well known, \cite{Kung}, such a term arises even in the hydrodynamic theory of equilibrium polar liquid crystals, as it should.}

{Indeed, the very mechanism invoked by the authors in other works \cite{solon2022susceptibility, dopierala2025inescapable} to anchor a rotational Nambu-Goldstone mode to lattice directions or imposed anisotropy would not work if the deterministic part of that mode were constrained to have a conservation law form. The mechanism \cite{Chaikin, Low_Manohar} through which the Nambu-Goldstone mode for rotations acquires a mass when translation symmetry is spontaneously broken would likewise fail too.} 

Furthermore, in \cite{Comment}, Chat\'e and Solon again claim that their simulations of the hydrodynamic equations of the immortal flock yield a mass. We discussed this in our original article, where we said (and we quote) ``Furthermore, {comparison with IV.6 suggests that}
their claimed numerical generation of a term $\propto\theta$, in the $\theta$ equation, by simulating Eq.~(22), (23) of their SI (see Fig.~S4 of \cite{Solon_Chate}) is a result of inadvertently breaking \emph{rotation} symmetry. Our analysis makes it clear that for Chat\'{e} and Solon's \cite{Solon_Chate} Eq.~(18) to be rotation invariant, their $h_{x,2}$ has to be equal to $\lambda_1$ (or equivalently, $h_{x,2} = -\tilde{\sigma}$ in (22), (23) in their SI); see Eq.~(III.5), and Eqs.~(IV.5) and (IV.6) of this article.  Indeed it is only for that case that their Fig.~S4 shows a massless correlator. Fig.~S4 shows that other choices of $h_{x,2}$ give a mass to their $\phi$ field, unsurprisingly, since they break rotation invariance regardless of the value of $h_{x1}$''.

We agree that Eq.~(IV.5-6) of \cite{Article}---which are also anisotropic---are explicitly symmetric under joint rotation of space and orientation, while the truncated ones are not. This is the case for many classical models; see, for instance,  \cite{Nelson_Pelcovits} in which {their} Eq.~(4.3) is explicitly invariant under the same transformation, but not Eq.~(4.4). The KPZ equation, which describe{s} the fluctuations of an interface moving normal to itself, is another example. This certainly doesn't imply that the dynamics of Goldstone modes should have a continuity equation form (indeed, the functional derivatives of the free energies in Eq.~(4.3) and (4.4) of \cite{Nelson_Pelcovits}---cited in our article---do not have this form)!
It does mean that if one wants to ensure that a simulation doesn't produce terms that are forbidden by symmetry (such as a mass term) one may be better off simulating the full equation that is explicitly invariant under the required symmetry, and not a truncated one. One {could} of course, use the truncated equation to analytically calculate renormalisation group flows, etc, {when the truncated terms are irrelevant}. 

Their discussion about the ``definition'' of hydrodynamic equations is misplaced. Indeed, if one uses the linear theory to calculate the relevance of nonlinearities (as is usually done), one finds that all nonlinearities in Eq.~(IV.5-6) are equally relevant.

Finally, while the deterministic part of our equation for the Goldstone mode in the Malthusian case is indeed expressable as a total divergence, that is an accident. This would not have been the case if we had retained nonlinear terms at $\mathcal{O}(\nabla^2)$. 

To summarise, we {have clarified once again} 
that there is no argument whatsoever for the dynamics of Goldstone modes to have a conservation law form. The article \cite{Solon_Chate} and the comment \cite{Comment} present no evidence for this claim; indeed, it \emph{can't} since the claim itself is {incorrect} {as shown by an enormous number of counterexamples.} 

\item We did not refer to \cite{CFLee} in our article because it is not directly relevant. This work examines a flocking model in the presence of a global constraint. The dynamics for the Goldstone mode of this model contains additional relevant nonlinearities that do not appear in unconstrained systems 
that we consider and lacks a nonlinear term that we show controls the hydrodynamics of unconstrained Malthusian flocks and unconstrained flocks whose hydrodynamic equations do not contain any nonlinearities involving the density (their simplification 2) in two dimensions.
The non-renormalisation of the noise strength there is ultimately a consequence of their simplification 1 and 2.

\item We, of course, agree that the nonlinearity $\propto\partial_x\theta^3$ is ``relevant in the RG sense''; {we disagree on why it is relevant and what the consequences of its relevance are. In particular, in contrast to \cite{Solon_Chate}, we show that rotation invariance implies that a fixed point at which the $\partial_y\theta^2$ nonlinearity is relevant, but the $\partial_x\theta^3$ nonlinearity is irrelevant, is unstable. We further discuss why an infinite number of other nonlinearities, which are naively all equally relevant, are actually irrelevant.
We also demonstrate that the two nonlinearities $\partial_y\theta^2$ and $\partial_x\theta^3$ \emph{should} have the same graphical correction (yielding a hyperscaling relation between them, in addition to the one that comes from the non-renormalisation of the noise strength) \emph{due to rotation invariance}. No equivalent discussion is presented by \cite{Solon_Chate} and \cite{Comment}. Instead, in their work, this is one among an infinite number of terms, all naively equally relevant, that arise from their Eq.~(6), which is obtained by incorrectly insisting that the deterministic part of the equation for Goldstone modes must have a conservation law form. They further don't explicitly argue that $\partial_y\theta^2$ and $\partial_x\theta^3$ nonlinearities must have the same graphical corrections (in particular, this is not required for retaining the conservation law form); instead, they incorrectly argue that neither should acquire \emph{any} graphical corrections.}

\item We do not understand the claims of \cite{Solon_Chate} and \cite{Comment} regarding the invariance of their dynamical equation{. In particular, we disagree with \cite{Comment} that the symmetry they propose is only broken at order $\theta_0^2$. Integrating their Eq.~(3) with respect to time, yields time-non-local contributions to linear order in $\theta_0$ in the transformations of $\partial_x \theta$ and $\partial_x \theta$ (see Eq.~(III.62) and (III.63) of \cite{Article}) that are not cancelled out. {Further, plugging Eqs. (III.61), (III.62) and (III.63) into (III.59), yields terms linear in $\theta_0$ that are absent in (III.58).}
We have also} not been able to find a{ny other} symmetry which forces the nonlinearities to remain unrenormalised and don't think such a symmetry exists. {We further note that \cite{Comment} and \cite{Solon_Chate} agree that their equation of motion for Malthusian flocks \emph{does not} actually have pseudo-Galilean invariance.}

\item We reiterate that we cannot calculate the exponents for either the immortal flock or the Malthusian flock analytically. Therefore, the exponents of the Malthusian flock may be arbitrarily close to the ones numerically obtained in \cite{Solon_Chate}, though we would be surprised if they were \emph{exactly} $\chi=-1/4$, $\zeta=3/4$, $z=5/4$ (since that would mean that nonlinearities receive no graphical correction and we have not been able to find any argument for it). It is also possible that nonlinearities receive no graphical correction at \emph{one}-loop order and therefore, one has to go to really large scales to see truly asymptotic behaviour. 

Further, since there is no argument that forces the noise strength in immortal flocks to remain graphically uncorrected, we generically expect it to acquire graphical corrections. Indeed, \emph{even} in Malthusian flocks in $d>2${\textemdash}where the non-stochastic part of the dynamical equation for the Goldstone mode cannot be written as a total divergence{\textemdash}the noise strength {acquires} graphical corrections \cite{Mal_RG1, Mal_RG2, John_bk}. It is possible that the graphical correction of noise strength for immortal flocks in $d=2$ is small and, therefore, not easily detectable numerically. It is also not impossible that the fixed point is controlled \emph{only} by a set of nonlinearities that doesn't allow the renormalisation of the noise strength (for instance, because nonlinearities that contain both density fluctuations $\delta\rho$ and angular fluctuations $\theta$ in the $\theta$ equation turn out to be irrelevant). We do not know which of these scenarios is correct; indeed, we cannot make \emph{any} analytical prediction about the exponents for the ordered phase of the immortal flock. However, we 
{emphasize once again} that this does not, in any way, imply that the equation for the Goldstone mode must have a conservation law form. Further, we do not find any justification for the non-renormalisation of nonlinear terms claimed in \cite{Solon_Chate}.
\item {The numerical results are \emph{not}   a compelling argument for the correctness of the analytic argument. To illustrate this, imagine that someone proposed that the coefficient  $c$ of the 
$c|\nabla\bf{M}|^2$ term in the Landau free energy of an $O(n)$ model was unrenormalized. We know, of course, that such an argument is incorrect, since the correction to that coefficient has been explicitly calculated in the $4-\epsilon$ expansion, and found to be non-zero.

Nonetheless, because the critical exponent $\eta$ associated with this renormalization proves to be very small ($\eta\sim.02)$, the \emph{true} scaling law $\nu=\gamma/(2-\eta)$ (where $\nu$ and $\gamma$ are respectively the correlation length and susceptibility critical exponents) agrees with the \emph{incorrect} scaling law $\nu=\gamma/2$, which one would propose if one erroneously believed that $c$ was unrenormalized, to 1\% accuracy, which is even better than the agreement between Chat\'e and Solon's numerical result and the prediction of their own incorrect argument. Such fortuitous agreement does not make their argument any more correct than the claim that $c $ is unrenormalized is.}

\end{itemize}


{We do not dispute the numerical results of \cite{Solon_Chate} or \cite{Mahault}
; rather, we disagree with the analytical claims made in \cite{Solon_Chate} and have shown them to be incorrect.}

	\bibliographystyle{apsrev4-2}
\bibliography{ref}
\end{document}